\documentstyle[psfig,epsfig]{mn}
\newif\ifAMStwofonts

\def\ah{\hbox{$^{\rm h}$}}
\def\am{\hbox{$^{\rm m}$}}

\def\Mpc{{\rm\thinspace Mpc}\,\,}

\def\kms{\hbox{km\,s$^{-1}$}\,\,}

\def\kmsmpc{\hbox{km\,s$^{-1}$\,\Mpc$^{-1}$}\,}
\def\ML{\hbox{$M_{HI}/L_{B}$}\,\,}

\def\msun{\hbox{${\rm\thinspace M}_{\odot}$}}

\def\Bj{\hbox{$B_J$}}

\def\sqd{\hbox{deg$^2$\,\,}}

\def\hi{\hbox{H{\sc{i}\,\,}}}
\def\HI{\hbox{H{\sc{i}\,\,}}}
\def\mhi{\hbox{$M_{HI}$}\,\,}
\def\hirich{\hbox{H{\sc{i}}}{-rich}\,\,}

\def\miriad{\hbox{\textsc{miriad}}\,\,}
\def\mbspect{\hbox{\textsc{mbspect}}\,\,}
\def\gfinder{\hbox{\textsc{multifind}}}

\ifoldfss
  \ifCUPmtlplainloaded \else
    \NewTextAlphabet{textbfit} {cmbxti10} {}
    \NewTextAlphabet{textbfss} {cmssbx10} {}
    \NewMathAlphabet{mathbfit} {cmbxti10} {} % for math mode
    \NewMathAlphabet{mathbfss} {cmssbx10} {} %  "   "    "
  \fi
  \ifAMStwofonts
    \ifCUPmtlplainloaded \else
      \NewSymbolFont{upmath} {eurm10}
      \NewSymbolFont{AMSa} {msam10}
      \NewMathSymbol{\upi}     {0}{upmath}{19}
      \NewMathSymbol{\umu}     {0}{upmath}{16}
      \NewMathSymbol{\upartial}{0}{upmath}{40}
      \NewMathSymbol{\leqslant}{3}{AMSa}{36}
      \NewMathSymbol{\geqslant}{3}{AMSa}{3E}

      \let\leq=\leqslant 
       
    \fi
  \fi
\fi % End of OFSS

\ifnfssone
  \newmathalphabet{\mathit}
  \addtoversion{normal}{\mathit}{cmr}{m}{it}
  \addtoversion{bold}{\mathit}{cmr}{bx}{it}
  \newmathalphabet{\mathbfit} % math mode version of \textbfit{..}
  \addtoversion{normal}{\mathbfit}{cmr}{bx}{it}
  \addtoversion{bold}{\mathbfit}{cmr}{bx}{it}
  \newmathalphabet{\mathbfss} % math mode version of \textbfss{..}
  \addtoversion{normal}{\mathbfss}{cmss}{bx}{n}
  \addtoversion{bold}{\mathbfss}{cmss}{bx}{n}
  \ifAMStwofonts
    \ifCUPmtlplainloaded \else
      \UseAMStwoboldmath
      \makeatletter
      \new@mathgroup\upmath@group
      \define@mathgroup\mv@normal\upmath@group{eur}{m}{n}
      \define@mathgroup\mv@bold\upmath@group{eur}{b}{n}
      \edef\UPM{\hexnumber\upmath@group}
      \new@mathgroup\amsa@group
      \define@mathgroup\mv@normal\amsa@group{msa}{m}{n}
      \define@mathgroup\mv@bold\amsa@group{msa}{m}{n}
      \edef\AMSa{\hexnumber\amsa@group}
      \makeatother
      \mathchardef\upi="0\UPM19
      \mathchardef\umu="0\UPM16
      \mathchardef\upartial="0\UPM40
      \mathchardef\leqslant="3\AMSa36
      \mathchardef\geqslant="3\AMSa3E

      \let\leq=\leqslant 

    \fi
  \fi
\fi % End of NFSS release 1

\ifnfsstwo
  \DeclareMathAlphabet{\mathbfit}{OT1}{cmr}{bx}{it}
  \SetMathAlphabet\mathbfit{bold}{OT1}{cmr}{bx}{it}
  \DeclareMathAlphabet{\mathbfss}{OT1}{cmss}{bx}{n}
  \SetMathAlphabet\mathbfss{bold}{OT1}{cmss}{bx}{n}
  \ifAMStwofonts
    \ifCUPmtlplainloaded \else
      \DeclareSymbolFont{UPM}{U}{eur}{m}{n}
      \SetSymbolFont{UPM}{bold}{U}{eur}{b}{n}
      \DeclareSymbolFont{AMSa}{U}{msa}{m}{n}
      \DeclareMathSymbol{\upi}{0}{UPM}{"19}
      \DeclareMathSymbol{\umu}{0}{UPM}{"16}
      \DeclareMathSymbol{\upartial}{0}{UPM}{"40}
      \DeclareMathSymbol{\leqslant}{3}{AMSa}{"36}
      \DeclareMathSymbol{\geqslant}{3}{AMSa}{"3E}

      \let\leq=\leqslant 

    \fi
  \fi
\fi % End of NFSS release 2

\ifCUPmtlplainloaded \else
  \ifAMStwofonts \else % If no AMS fonts
    \def\upi{\pi}
    \def\umu{\mu}
    \def\upartial{\partial}
  \fi
\fi

\title[HI in the Fornax Region]
  {The Large Scale Distribution of Neutral Hydrogen in the Fornax Region}
\author[M. Waugh et al.]
  {Meryl Waugh$^1$, M.J. Drinkwater$^1$, R.L. Webster$^1$, 
L. Staveley-Smith$^2$, 
\newauthor V.A. Kilborn$^3$, D.G. Barnes$^4$, R. Bhathal$^5$,   
W.J.G. de Blok$^2$,  P.J. Boyce$^6$, 
\newauthor M.J. Disney$^6$, R.D. Ekers$^2$,  K.C. Freeman$^7$, B.K. Gibson$^4$,
P.A.  Henning$^8$, H. Jerjen$^7$, 
\newauthor P.M. Knezek$^9$, B. Koribalski$^2$,  M. Marquarding$^2$, 
R.F. Minchin$^6$, R.M. Price$^8$,   
\newauthor M.E. Putman$^{10}$, S.D. Ryder$^{11}$, E.M. Sadler$^{12}$,  
F. Stootman$^5$ and M.A. Zwaan$^1$\\ 
    $^1$School of Physics, University of Melbourne, VIC  3010, Australia\\
    $^2$Australia Telescope National Facility, CSIRO, P.O. Box 76, Epping, NSW
          1710, Australia\\
    $^3$University of Manchester, Jodrell Bank Observatory, Macclesfield, 
          Cheshire, UK, SKll 9DL\\
    $^4$Centre for Astrophysics \& Supercomputing, Swinburne University, 
          Mail \#31, P.O. Box 218, Hawthorn, VIC  3122, Australia\\
    $^5$University of Western Sydney Macarthur, Department of Physics, 
          P.O. Box 555, Campbelltown, NSW 2560, Australia\\
    $^6$University of Wales, Cardiff, Department of Physics \& Astronomy, 
          P.O. Box 913, Cardiff CF2 3YB, U.K.\\
    $^7$Research School of Astronomy and Astrophysics, 
          ANU, Weston Creek P.O., Weston, ACT 2611, Australia\\
    $^8$University of New Mexico, Department of Physics \& 
          Astronomy, 800 Yale Blvd. NE, Albuquerque, NM 87131, USA\\
    $^9$Space Telescope Science Institute, 3700 San Martin
          Drive, Baltimore, MD, 21218, USA\\
    $^{10}$Center for Astrophysics and Space Astronomy, University of Colorado,
 Boulder, CO 80309-0389, USA. Hubble Fellow\\
    $^{11}$Anglo-Australian Observatory, P.O. Box 296, 
          Epping, NSW 1710,  Australia\\
    $^{12}$University of Sydney, Astrophysics Department, School 
           of Physics, A28, Sydney, NSW 2006, Australia\\
}

\date{Accepted 0000 Month 00. Received 0000 Month 00}

\pagerange{\pageref{firstpage}--\pageref{lastpage}}
\pubyear{0000}

\begin{document}
 
\maketitle

% -------------  ABSTRACT -----------

\label{firstpage} 

\begin{abstract}  

Using data from the \hi Parkes All Sky Survey (HIPASS), we have
searched for neutral hydrogen in galaxies in a region 
$\sim$25$\times$25 \sqd centred on NGC\,1399, the
nominal centre of the Fornax cluster.  Within a velocity search range
of 300--3700 \kms and to a 3$\sigma$ lower flux limit of $\sim$40 mJy, 
110 galaxies with \hi emission were detected,  one of which is 
previously uncatalogued.  None of the detections has early-type
morphology. Previously unknown velocities for 14 galaxies have been 
determined, with 
a further 4 velocity measurements being significantly dissimilar to
published values.  Identification of an optical counterpart 
is relatively unambiguous for more than $\sim$90 per cent  of our
\hi galaxies. 
The galaxies appear to be embedded in a sheet at the cluster velocity
which extends for more than 30\degr\ across the search area.  At the
nominal cluster distance of $\sim$20 Mpc, this corresponds to an
elongated structure more than 10 Mpc in extent. A velocity gradient
across the structure is detected, with radial velocities increasing by  
$\sim$500 \kms from SE to NW.  The clustering of galaxies evident in optical 
surveys is only weakly
suggested in the spatial distribution of our \hi\
detections. Of 62 \hi detections within a 10\degr\ projected radius of
the cluster centre, only two are within the core region (projected radius
$<$1\degr) and less than 30 per cent  are within 3.5\degr, suggesting a
considerable deficit of \hirich\ galaxies in the centre of the
cluster.  However, relative to the field, there is a 3($\pm$1)-fold
excess of \hirich\ galaxies in the outer parts of the cluster where
galaxies may be infalling towards the cluster for the first time.

\end{abstract}


%------------  section INTRODUCTION   -----------

\section{INTRODUCTION --- THE FORNAX CLUSTER}
\label{sec_intro} 

The Fornax Cluster is amongst the closest and most well studied clusters 
in the southern sky, providing an interesting nearby field for the study 
of large scale structure (LSS), galaxy populations, dynamics and evolution 
in the cluster environment.
Although the cluster is not as rich as other well-studied clusters, its 
core galaxy density is relatively high (Ferguson 1989a).

The central galaxy of the Fornax Cluster is the giant elliptical NGC\,1399  
with a radial velocity of $\sim$1425 \kms.  Its position, 
R.A.\  3\ah38\am29\fs3, Dec.\  $-$35\degr27\arcmin01\arcsec\  
(J2000), is generally 
adopted as the  approximate centre of the cluster.  Also particularly 
well-studied are the barred spiral NGC\,1365 and the strong radio continuum 
source NGC\,1316 {\mbox{(Fornax A)}}.  Although the Cepheid distances to 
three Fornax 
spirals have been determined (NGC\,1365 at 18.6 $\pm$ 0.6\Mpc, 
NGC\,1326A at 18.7 $\pm$ 2.0\Mpc\  and 
NGC\,1425 at 22.2 $\pm$ 2.7\Mpc) (Madore et al.\ 1999, Prosser et al.\ 1999 and  
Mould et al.\ 2000, respectively), an accurate cluster distance is still
somewhat uncertain, particularly in the light of the substructure analysis of 
Drinkwater et al.\ (2001a).  In this paper, we adopt a value close to the mean 
of the three Cepheid estimates, 20 Mpc, as the cluster distance.

Recessional velocities of Fornax Cluster galaxies range from about 700 to 2200 
\kms (Drinkwater et al.\ 2001b).  
No galaxies have been detected, either optically 
or in {H{\sc{i}}, in front of the cluster and few in the void behind it, in 
the region 2200--4500 \kms. Galaxies beyond about $cz = 3000$ \kms would not be 
classified as likely cluster members.  

Covering an area of nearly 40 deg$^2$, Ferguson's optical Fornax
Cluster Catalogue (FCC) (Ferguson 1989b) identified 340 likely cluster
members, only 85 of which had published redshifts.  Subsequent surveys
have reassigned some members from likely membership to background
status and \textit{vice versa} and have also determined some
previously unknown redshifts (e.g.\ Drinkwater \& Gregg 1998, Drinkwater et al.\ 2001b, Schr\"{o}der, Drinkwater \& Richter 2001).

Drinkwater et al.\ (2001b) have  recently measured the redshifts of more than 
$500$ bright galaxies ($16.5<\Bj<18$) in a 6$\times$6 \sqd\ field 
centred on Fornax and using the FLAIR-II spectrograph on the UK Schmidt 
Telescope.  A deeper unbiased Fornax Cluster Spectroscopic Survey (FCSS) is 
targetting {\it{all}} objects ($\sim14000$) within the magnitude limits 
16.5 $\leq\Bj\leq$ 19.7 in four $\pi$ deg$^2$ fields 
(total area $\sim$12 deg$^2$), utilising  
the 2 degree Field (2dF) spectrograph of the Anglo-Australian Telescope 
(Drinkwater et al.\ 2000).

Several pointed neutral hydrogen surveys of Fornax galaxies have been
reported (e.g.\ Schr\"{o}der et al.\ 2001, Bureau, Mould \& Staveley-Smith 1996).  
The first blind \HI survey of Fornax was conducted by Barnes et al.\
(1997) and covered about 8\degr$\times$8\degr\ of the central cluster
region.  We report here on a larger, deeper, blind survey of the
region, with detection rates about three times greater.

Utilising the 13-beam receiver on the Parkes 64 m radio telescope,
 the \HI Parkes All Sky Survey (HIPASS) covered the entire southern
 sky south of $\delta = 2$\degr\ for $\lambda\ 21$cm neutral hydrogen
 (Staveley-Smith et al.\ 1996). Observations were completed in March
 2000 and this ambitious project has enormous scientific potential,
 providing a wealth of new \HI data, unbiased by optical selection
 criteria or effects.  For this paper, we sampled about 3 per cent  of the
 HIPASS dataset (13 of 388 cubes) to examine the distribution, number
 density and \HI content of galaxies over a 25$\times$25 \sqd\
 field centred on the Fornax cluster but extending well beyond it. Our
 results contribute new information pertaining to galaxy evolution and
 dynamics and the large scale structure (LSS) of the local Universe in
 the Fornax region.

The fields covered by several surveys of the Fornax cluster are illustrated 
in Fig.~\ref{fig_surveys}. Also shown are the positions of the 13 HIPASS 
cubes used to compile our mosaic.

Further details and parameters of HIPASS are 
outlined in Section~\ref{sec_data}.  Galaxies were detected and confirmed in 
the mosaic by the methods described in Section~\ref{sec_detects}.
 The properties of the galaxies are examined in Section~\ref{sec_disc} 
and the results of this study are summarised in Section~\ref{sec_concl}.

%----------- fig 1  -----------
\begin{figure}
\begin{center}
\psfig{file=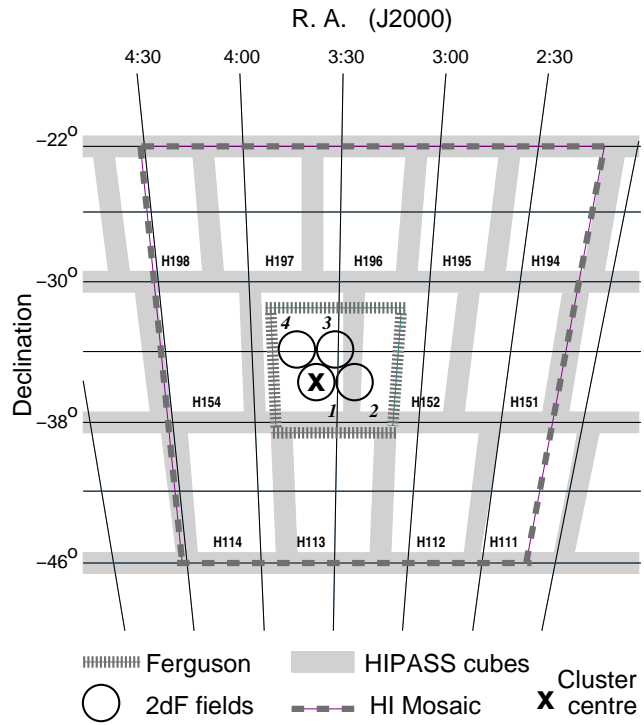,angle=0,width=85mm}
\caption{Relative positions and approximate sizes of various optical and 
\hi surveys in the Fornax region. The Ferguson Cluster Catalog region is 
indicated by the thin dashed lines, the 2dF survey fields are shown as 
circles and the approximate positions of standard HIPASS cubes are outlined. 
The area covered by our Fornax \hi mosaic is indicated by heavy dashed 
lines. The cross shows the position of NGC\,1399 at the cluster centre.}
\label{fig_surveys}
\end{center}
\end{figure}

%----------- section DATA SAMPLE -----------
\section{DATA SAMPLE} 
\label{sec_data}

The characteristics of the Parkes multibeam system used for HIPASS are 
described by Staveley-Smith et al.\ (1996). Barnes et al.\ (2001) have 
comprehensively detailed the survey observing parameters, data collection, 
calibration and imaging techniques. 
For the survey, the telescope scanned in declination strips of 8\degr, 
each 1.7\degr\ wide. The scanning procedure was optimized to ensure 
approximately 
uniform sky coverage at full sensitivity whilst still allowing online data 
processing to be automated and statistically robust.  The mean beamwidth is 
14.3 arcmin full width at half-maximum (FWHM) and the total integration time is 460 s beam$^{-1}$.  After 
median gridding, data were processed into cubes measuring 
8$\times$8 \sqd\ on the sky, with overlap of up to 
$\sim$1\degr. Pixel size is 4 arcminutes. The survey has a bandwidth of 
64 MHz, detecting \HI emission 
in the range $-1280<cz<12700$ \kms over 1024 planes at a separation of 
13.2 km\,s$^{-1}$. The velocity resolution after smoothing is 
18 km\,s$^{-1}$.  

The final HIPASS data have an rms noise level of around 13.3 mJy and
thus a 3$\sigma$ detection limit of $\sim$40 mJy beam$^{-1}$.  The gridded
HIPASS beam has a FWHM of 15.5 arcmin and an image pixel size of 4
arcmin.  Most sources in our mosaic survey are essentially unresolved
and a final positional accuracy of 3--4\arcmin\ is expected, depending
on the flux (Barnes  1998).  HIPASS detects \HI masses to a 3$\sigma$
lower limit of {\mbox{$\sim$$10^6$$ d^2$\msun}}\footnote{for an \HI
linewidth of 100 km\,s$^{-1}$, where $d$ is the distance to the source in
\Mpc\ and $H_0 = 75$ \kmsmpc}, corresponding to $\sim$4$\times10^8
\msun$ at the distance of Fornax.

For this Fornax region survey, 13  HIPASS data cubes were mosaicked together 
before analysis and trimmed to create a single, large,  
$\sim$25$\times$25 \sqd\ datacube. Mosaicking the cubes improves the 
effective rms noise levels at the positions of the HIPASS cube edges and 
creates a 
single dataset with relatively smooth coverage across the entire field.

The projected sky area covered by the mosaic is represented in 
Fig.~\ref{fig_surveys}.  An image of the data, Fig.~\ref{fig_mom}, shows 
galaxies in peak flux as bright bands, the length of each being indicative of
 the 
velocity width of the detections. In this velocity versus RA map, the 
cluster galaxies  appear to be embedded in a long sheet of galaxies at similar
velocities which crosses  the mosaic region. The void behind the cluster is 
clearly seen.  The detection at approximately 2900 km\,s$^{-1}$, 3\ah08\am, is at a 
declination of $-$23\degr\ and is not near the cluster.

%-----------  section  GALAXY DETECTIONS -----------
\section{H{\sc{i}\,\,}  GALAXY DETECTIONS} 
\label{sec_detects}

\subsection{Some HIPASS characteristics}
\label{sec_charact}

Our survey is limited by two particular selection effects.  First we
find a clear lower cutoff in peak flux (at $\sim$30 mJy), reflecting
the galaxy search technique. Secondly, there is a cutoff in velocity
width (at $\sim$50 \kms) due to the spectral resolution of the survey.

Sources with extended continuum emission can produce standing wave
patterns. After bandpass removal a ripple remains in the spectrum,
making these sources difficult to parametrize. For example and in
particular, we were unable to characterize NGC\,1316 (the strong radio
source Fornax A) in our data.  Spectra of sources within a few
arcminutes of the continuum source are also affected by the ripple. In
the vicinity of bright continuum sources, the online bandpass removal
process distorts the spectral baseline at both the low and high
frequency ends, making the detection of nearby \HI sources more
difficult.  The post-processing application of a cleaning algorithm to
each cube has improved the bandpass removal and the detection rate
close to continuum sources (Barnes et al.\ 2001).

\subsection{Automated galaxy finder}
\label{subsec_GF}

After masking  the edges of the mosaic, an automated galaxy finding 
algorithm, \gfinder,  developed by Kilborn (2001), was used to search for 
candidate detections.  

The galaxy finder searches the data in 3 dimensions, (RA, Dec. and
radial velocity), using \miriad routines (Sault, Teuben \& Wright 1995) to
perform the calculations.  In concentrating our search on the Fornax
region, the finder was set to search the mosaic in the range 300--3700
\kms to a lower peak flux limit of $3\sigma$, corresponding on average
to about 40 mJy.  In \gfinder, each pixel is searched in velocity
space, plane by plane.  Any detections above the preset signal-to-noise ratio 
(S/N) cutoff  
and in two or more consecutive velocity planes are
retained as possible detections.  The search was repeated after each
of two rounds of Hanning
smoothing in the velocity planes --- this type of binning can reduce
the rms noise of the baseline, effectively increasing the S/N, and has
the potential to detect extra sources with low S/N profiles. The
finder fits a Gaussian to each candidate detection to estimate the
central position (although HIPASS sources are not necessarily
Gaussian) and uses \mbspect in \miriad to determine spectral parameters.

A final candiate list of over 800 \hi  detections was generated, with 
estimates  of RA, Dec., velocity,
velocity width, peak flux and the smoothing level at which the detection was 
first made. 

%----------- fig 2 -----------
\begin{figure}
\psfig{file=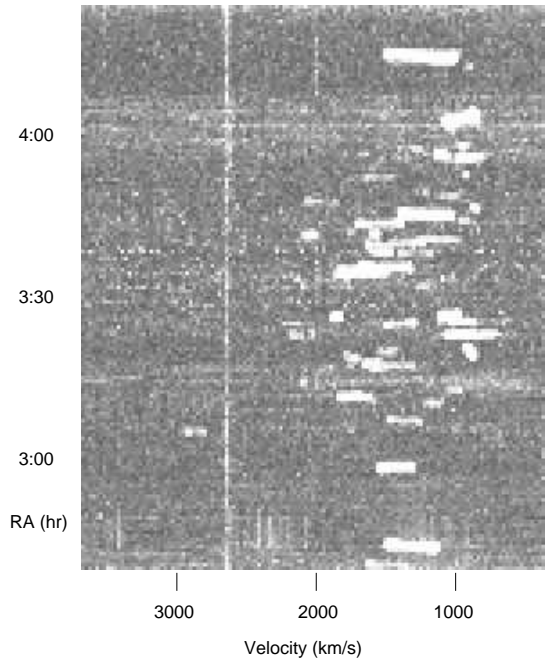,angle=0,width=90mm}
\centering
\caption{HI flux map from the Fornax mosaic data showing velocities,
velocity widths and RA.  Detections appear as bright bands in this
{\sc{kview}} image.  The interference band around 2630 \kms\ (1408
MHz) is seen as the bright line present through all RA positions. Note
that velocities here are heliocentric, without the Local Group
correction (see Section~\ref{subsec_calcs_ML}).}
\label{fig_mom}
\end{figure}

\subsection{Confirmed detections}
\label{subsec_conf} 

The preliminary detection list was first culled by removing the unlikely 
detections
at  2630 $\pm$ 20 \kms\ which are known to result from telescope correlator 
interference (see Fig.\ 2). To reduce positional redundancies,  
candidates listed within 2 pixels of each other were grouped together.  
This left a list of more than 400 to be checked individually.  
Each candidate detection was visually inspected 
and parameterized using {\sc{\miriad}} software packages. 
An outline of the steps follows:

\begin{enumerate}
  \item A  moment map of flux integrated over the velocity planes in which 
the candidate was detected was made. 

  \item A Gaussian was fitted to the moment map and the 
central positon calculated.

  \item A spectrum of \hi emission at this position was 
generated and a Guassian-smoothed baseline fitted. Spectral parameters 
were calculated from unsmoothed spectra: 
heliocentric recessional velocity (at the midpoint of the 50 and 20 per cent  velocity 
widths),  velocity widths (at the 50 and 20 per cent  peak flux levels), peak 
flux and total integrated flux. For the majority of sources, measurements 
were based on the \miriad task \mbspect using the ``point'' setting. For 
extended sources (those so identified in Table 1), the input parameter ``sum'' 
was used. 

\end{enumerate}

After grouping duplicates, rejecting those unlikely detections which could 
not be successfully parameterized and those with spectra corrupted by 
continuum ripple (notably Fornax A and nearby sources), 110 \hi detections 
were confirmed to a maximum velocity cut-off limit of 3700 \kms. The
positions and measured parameters of the detections are listed in  Table 1.
(Notes on the table can be found following section ~\ref{sec_concl}.) The spectra are plotted in the Appendix.

The positions of the \hi detections relative to the position of the
cluster centre are shown in Fig.~\ref{fig_detects}.  Plots of the
velocity versus RA and Dec. are shown in Figs.~\ref{fig_conera}
and ~\ref{fig_conedec} respectively.  There are no detections in front
of the cluster and relatively few beyond 2500 \kms.  These plots show
clear evidence of void regions. The distribution of galaxies is
discussed further in Section ~\ref{subsec_V_LSS}.

%----------- fig 3  -----------
\begin{figure}   
\psfig{file=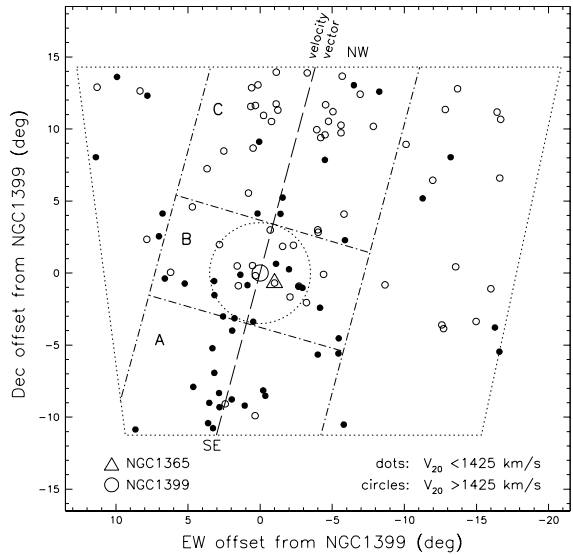,angle=0,width=80mm}
\centering
\caption{Positions of Fornax mosaic HIPASS detections. Detections at
$<$1425 \kms \ are shown as solid dots, those at $>$1425 \kms \ are shown as
open circles. The positions of NGC\,1399 (a giant elliptical, not 
detected) at the cluster centre and NGC\,1365 (a large spiral)
are shown. The dotted circle indicates the approximate cluster
boundary at a radial distance of 3.5\degr\ from NGC\,1399.  A velocity
gradient across the structure from SE to NW is suggested, with an
approximate angle as indicated by the velocity vector line (long
dashes).  The detections in the regions A, B \& C (dot-dash lines),
relative to the vector, are compared in Section~\ref{subsec_V_LSS}.}
\label{fig_detects}
\end{figure}

%----------- fig 4 -----------
\begin{figure}
\vspace{-1.0cm}
\psfig{file=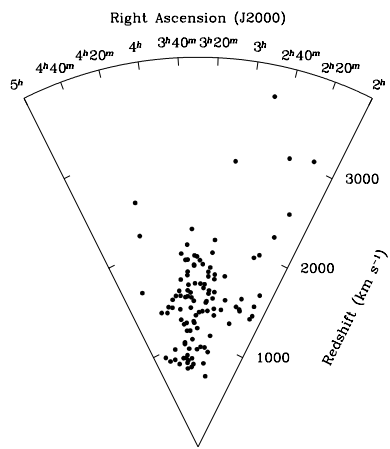,angle=0,width=85mm}
\centering
\vspace{-0.5cm}
\caption{\hi detections: heliocentric velocity vs. RA}
\label{fig_conera}
\end{figure}

%----------- fig 5  -----------
\begin{figure}
\vspace{-1.0cm}
\psfig{file=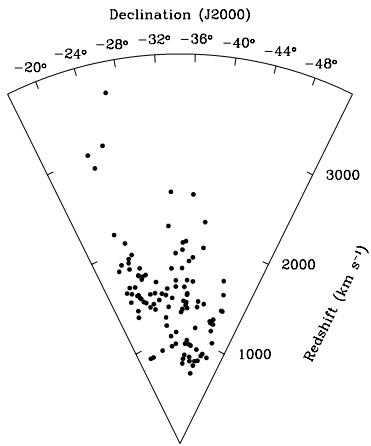,angle=0,width=85mm}
\centering
\vspace{-0.5cm}
\caption{\hi detections: heliocentric velocity vs. declination}
\label{fig_conedec}
\end{figure}

\subsection{Optical Counterparts of \hi Detections}
\label{subsec_opt}

\subsubsection{Optical Identifications}

The NASA/IPAC Extragalactic Database\footnote{This research has
made use of the NASA/IPAC Extragalactic Database (NED) which is
operated by the Jet Propulsion Laboratory, California Institute of
Technology, under contract with the National Aeronautics and Space
Administration.  Data used in this paper were current in 2001 December.} 
(NED) and the Digitized Sky Survey\footnote{The Digitized Sky
Survey (DSS), provided by the Space Telescope Science Institute is
based on photographic data from the UK Schmidt Telescope.} (DSS)  were
searched for likely optical counterparts within a maximum 15
arcmin radius.  Although positional matches were expected to be
within $\sim$3--4 arcmin (Barnes 1998), we searched to larger radii
looking for any galaxies which may be associated with the \hi
detection.

The majority of our detections could be matched to a
single optically catalogued counterpart with a published coincident
velocity.  For 14 objects, a close positional match was found with a
single optical galaxy which had no previously published velocity.  In
at least $\sim$20 per cent  of cases, more than one optical candidate was
visible in the DSS images, although in about half of these one of the
optical galaxies had a known velocity consistent with the
\hi object being identified. For four galaxies, the published velocity
(`$V_{\rm NED}$', column 13, Table 1) is significantly different from our
results (columns 4 and 5). A remark in column 16 (`V diff') identifies these
detections.

 Where the optical match to an \hi detection remains uncertain or
unknown, it has been marked `C' (for `confused') in column 16 of
Table 1.  Where a possible optical identification of confused sources
is known, it is also noted in column 16. One of our detections, J0317-24, 
is previously uncatalogued.  The optical counterpart of
our \hi detection J0317$-$37 is FCCB0035, a galaxy listed as
`background' in the Fornax Cluster Catalog . However, we measure a
heliocentric velocity of 981 km\,s$^{-1}$, indicating that it is more probably
a cluster member.

All of these confused or uncertain detections will require
follow-up observations at higher resolution to confirm their natures
and identities.

A histogram of the projected distance from each detection to its most likely 
optical counterpart is shown in Fig.~\ref{fig_posn_histo}.  The majority of 
matches are within 
the expected positional accuracy of 3--4 arcmin, with $\sim$86 per cent  being matched 
within 3 arcmin and $\sim$94 per cent  matching within 4 arcmin.
Our confirmed detections and their likely matches to 
previously  catalogued galaxies are detailed in Table 1.

\subsubsection{Morphological types}

Examination of the optical morphologies of our \hi detections, as
published in NED and as determined from DSS images, revealed that all
of our detections were late-type galaxies (or of uncertain
morphology).  This correlation of morphological type, \hi content and
proximity to the core of a cluster is discussed further in
Section~\ref{sec_disc}.

%----------- fig 6 -----------
\begin{figure}
\psfig{file=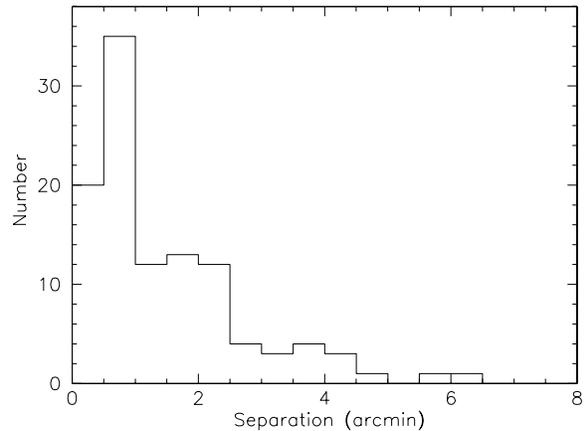,angle=-90,width=80mm}
\centering
\caption{Histogram of positional matches of each \hi detection with its 
most likely optical counterpart. Over 94 per cent  of matches are within 4 arcminutes.}
\label{fig_posn_histo}
\end{figure}

\subsection{Deriving H{\sc{i}\,\,}  Masses and mass-to-light ratios}
\label{subsec_calcs_ML}

The \hi mass of each detection was calculated using the standard 
formula (Giovanelli \& Haynes 1988), assuming detections to be 
unresolved in the Parkes telescope beam,

\begin{equation}
\mhi = 2.356\times 10^5 \, d^2 \, S_{int}
\end{equation}

\noindent where \mhi is \HI mass in units of \msun,  $d$  is distance in \Mpc\ 
 and  S{$_{int}$} is total integrated flux in Jy km\,s$^{-1}$, measured from the 
unsmoothed profiles. \newline

In deriving the \hi masses, detections were assumed to be either part
of the cluster or in the field. Detections within 3.5\degr\ of the
cluster centre were all assumed to be at the nominal cluster distance
of 20 Mpc. For galaxies more than 3.5\degr\ from the cluster centre
(i.e. assumed to be field galaxies), the distance was estimated from
the Hubble flow. The velocity of each field galaxy was measured from
the unsmoothed \hi profile at the mid-point of the 20 per cent  peak flux
level.  A Local Group correction was applied, using the IAU
convention,

\begin{equation}
V_{LG} = V + 300\sin\textit{l}\cos\textit{b}
\end{equation}

\noindent where $V_{LG}$ is the corrected velocity in km\,s$^{-1}$,  $V$ is 
the 
heliocentric velocity in km\,s$^{-1}$,  $\textit{l}$  is Galactic longitude 
and 
 $\textit{b}$ is Galactic latitude. \newline

Distances to field galaxies were then estimated from the Hubble equation,

\begin{equation}
d = V_{LG}/H_0
\end{equation}

\noindent where $d$ is distance in Mpc.  For these calculations, a
`Fornax-consistent' Hubble constant was calibrated such that at the
position of the cluster centre, the central cluster galaxy (NGC\,1399,
\textit{l} = 236.72\degr, \textit{b} = $-$53.63\degr, $V_{\rm hel}$ =
1425 km\,s$^{-1}$, $V_{LG}$ = 1276 km\,s$^{-1}$) has a distance of 20 \Mpc\ in the
Hubble equation. The value thus derived and used here is $H_0 = 64$
\kms Mpc$^{-1}$.

NED  was searched for apparent blue magnitudes of optical
counterparts so that, where possible, \ML values could be estimated
for the \hi detections.  When available, the apparent blue magnitudes
$m_{B}$, $B_T$ or $b_j$ were recorded and used in calculations of
luminosity. The value used and the magnitude type for each detection
are indicated in columns 10 and 11 of Table 1.

%----------- section DISCUSSION  -----------
\section{DISCUSSION}
\label{sec_disc}

\subsection{Projected number densities of detections}
\label{subsec_numdens}    

Greyscale surface number density plots of optical galaxies and
positions of \hi detections in the same redshift range are illustrated
in Fig.~\ref{fig_dens}.  The optical galaxies selected for this
comparison include only those in NED with published velocities in the
range 500--3700 km\,s$^{-1}$.  The ellipses refer to optical substructure
regions identified by Drinkwater et. al.\ (2001a), the centre of the
optical cluster being at the centre of the larger ellipse.  The
optical subcluster identified by the smaller ellipse is not distinguished
in {\hbox{H{\sc{i}}}.  The \hi emission appears to be relatively more
extended than the optical cluster region and more structure is suggested
by the apparent \hi overdensities in regions $\sim$10\degr\ to the
north-west (NW) and south-east (SE) of the cluster. The SE \hi
concentration, in particular, has a similar \hi content to the Fornax
cluster without the optical galaxy overdensity.

The surface number densities of \hi detections over the mosaic region are
shown in Fig.~\ref{fig_numdens}.  Our results suggest that the cluster
is detectable as an approximately three-fold overdensity relative to the
field beyond $\sim$4\degr\ from the cluster's optical centre. Within
4\degr\ ($\sim$1.4 Mpc), the number density of \hi detections is
0.46 $\pm$ 0.04 deg$^{-2}$ compared with 0.16 $\pm$ 0.03 deg$^{-2}$ in the
surrounding field (4\degr\ -- 10\degr\ from the centre).  Although
the number density of optical galaxies is seen to decline rapidly
beyond the nominal cluster boundary, the \hi density remains elevated
out to radii comparable with the cluster turnaround radius of
2.4\Mpc\ ($\sim$7\degr) (Drinkwater et al.\ 2001a).
At these radial distances,  galaxies are possibly 
infalling towards the cluster for the  first time and have presumably not 
yet been stripped of their gas (see Bekki, Couch \& Shioya 2002).

Very few \hi galaxies are found near the core  
of the cluster ($<$1\degr), and so a statistically significant galaxy 
density cannot be calculated for this region. In contrast, for galaxies 
listed in NED with velocities in the range 300--3700 
km\,s$^{-1}$, number densities are $\sim$45-fold greater in the core than at 
3--4\degr\ distant.

HIPASS is biased towards the detection of \hirich objects. As is
evident in Figure 3, the Fornax cluster of galaxies at the centre of
the mosaic field is not clearly distinguished in this survey. Of 65
\hi detections within a 10\degr\ projected radius of the cluster
centre, less than 30 per cent  are within the nominal optical cluster region
(projected radius $\sim3.5$\degr) --- an overdensity of 
$<$2$\frac{1}{2}$-fold 
compared to the number expected for an even distribution. Where the 
morphology is known, our
\hi detections are only late-type, a finding that is tightly
correlated with the very low \hi detection rate close to the cluster
core and with the well-established density-morphology relation 
(Dressler 1980).

Our results clearly indicate that, unlike optical searches, large
region \hi surveys such as HIPASS will not preferentially detect the
core region of galaxy clusters, although a modest overdensity of
{\hbox{H{\sc{i}}}{--rich}}/late-type galaxies may be seen in the
cluster infall regions beyond the core.

%----------- fig 7 -----------
\begin{figure}
\psfig{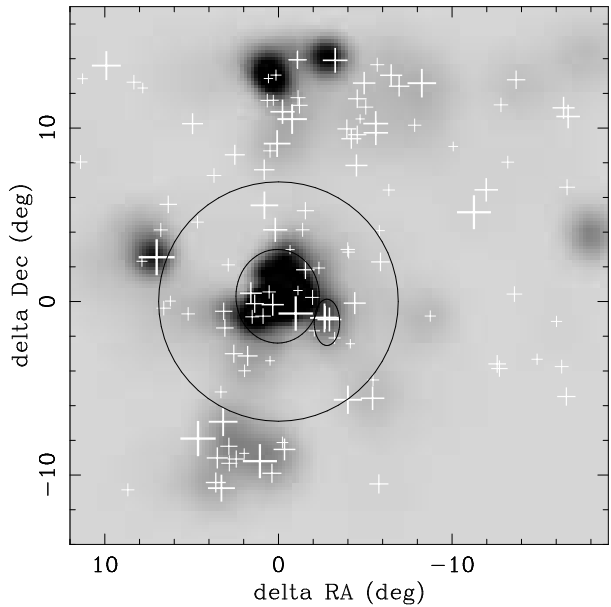}
\centering
\caption{Comparison of the distribution of HI-selected galaxies with
optically selected galaxies in the Fornax region. The grey scale shows
an adaptively smoothed measure of the number density of
optically selected galaxies. The white crosses indicate the positions
of all galaxies in the \hi sample with the size of the crosses
proportional to the logarithm of the integrated \hi flux. The two
ellipses show the $2\sigma$ limits of the main cluster and subcluster
and the large circle is at the cluster turnaround radius of
$\sim$7\degr\ as defined by the optical sample of Drinkwater et al.\
(2001a).}
\label{fig_dens}
\end{figure}

%----------- fig 8  -----------
\begin{figure}
\psfig{file=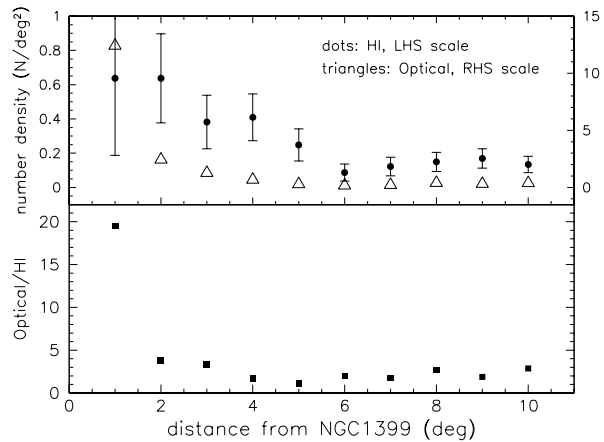,angle=-90,width=80mm}
\centering
\caption{The top panel shows the projected number density of detections at
increasing radii from the centre of Fornax. Full circles and the left-hand 
 scale are \hi detections; open triangles and the right-hand scale
are for optical counts from NED. Poisson number errors are shown for
the \hi detections.  The lower panel shows the ratio of
number densities: optically catalogued$/$\hi.}
\label{fig_numdens}
\end{figure}

\subsection{Velocity Measurements and Large Scale Structure}
\label{subsec_V_LSS} 

Velocity histograms for $\sim$200 optical galaxies listed by NED with
published redshifts in the Fornax mosaic area and for 110 \hi
detections in our mosaic are plotted in
Fig.~\ref{fig_histoV}. Galaxies within 3.5\degr\ of the cluster centre
are compared with those field galaxies further than 3.5\degr\ from
NGC\,1399. The cluster is well defined in velocity, with no objects in
front of the cluster and none catalogued between 2200  and 2800 \kms
in the void behind the cluster. This histogram shows that the
large-scale velocity distribution is similar to the cluster velocity
distribution.  Thus, the large-scale structure (LSS) is confined to a
narrow plane almost perpendicular to the line of sight.

The mean heliocentric velocity of the cluster \hi galaxies is 1516
$\pm$ 95 \kms, in agreement with the optical results of 
Drinkwater et
al.\ (2001a).  The velocity dispersion ($\sigma_V$) of these cluster
galaxies is 425 $\pm$ 71 km\,s$^{-1}$ --- considerably higher than for the
giant cluster galaxies of Drinkwater's optical sample ($\sigma_V = 308
\pm 30$ km\,s$^{-1}$) and thus indicative of an infalling population.

%----------- fig 9  -----------
\begin{figure}
\psfig{file=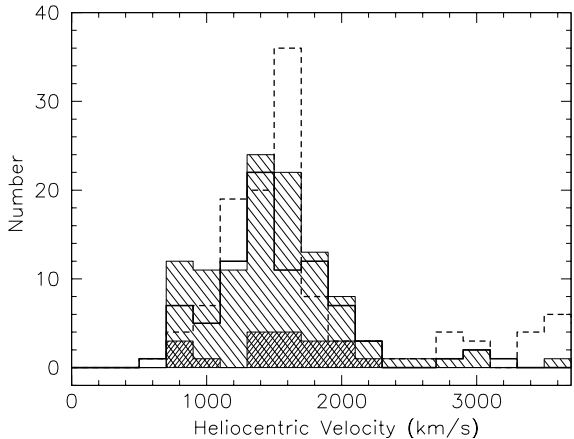,angle=-90,width=80mm}
\centering
\caption{Velocity histograms for Fornax Region galaxies: \hi
 detections within 3.5\degr\ of the cluster centre (heavy shading),
 \hi  detections beyond 3.5\degr\ (light shading), optical
 cluster galaxies (solid line), optical field galaxies (dashed line).
 The void in front of and behind the cluster beyond $\sim$2200 \kms is
 evident for both \hi and optical galaxies.}
\label{fig_histoV}
\end{figure}

%----------- fig 10 -----------
\begin{figure}
\psfig{file=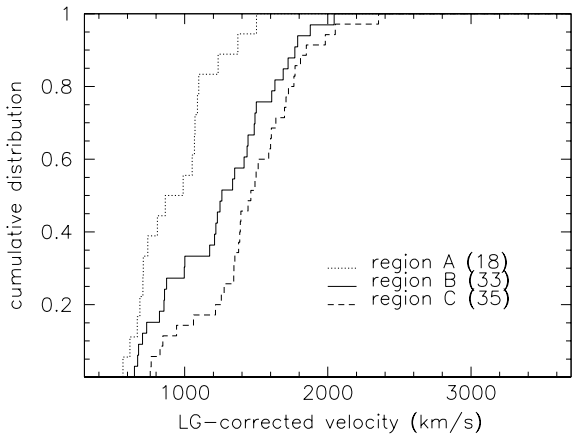,angle=-90,width=80mm}
\centering
\caption{The cumulative distribution function of Fornax region galaxy
velocities.  The three curves correspond to the regions A, B and C
indicated in Fig.~\ref{fig_detects} and measure LG-corrected
velocities. Galaxy numbers in each region are 18, 33 and 35, respectively.}
\label{fig_V_cdf}
\end{figure}

%----------- fig 11 -----------
\begin{figure}
\psfig{file=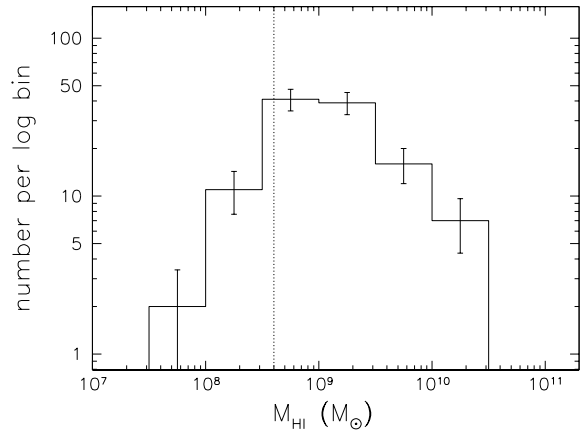,angle=-90,width=80mm}
\centering
\caption{The \hi masses of Fornax Region galaxies.  Error bars are calculated 
from number statistics ($\sqrt{N}$). At the Fornax distance, HIPASS detection 
rates become incomplete below $\sim4\times10^8 \msun$ (dotted line).}
\label{fig_HIMF}
\end{figure}

In Fig.~\ref{fig_detects}, the positions of the \hi detections
relative to the Fornax centre are shown.  Large-scale structure is
also evident in Fig.~\ref{fig_mom}, where the galaxies appear to be in
a long sheet-like arrangement in which the velocity widths of the
galaxies are comparable to the velocity dispersion of the whole
sheet. 

Our results indicate that there is a measurable velocity gradient,
from SE to NW, across the formation.  A vector at a sky angle of
$\sim$75\degr\ shows the direction of maximum velocity bias across the
region (see Fig.~\ref{fig_detects}).  The cumulative distribution
function of the velocity of \hi detections is shown in
Fig.~\ref{fig_V_cdf}, comparing the regions A, B and C. The three
regions are cut relative to the velocity vector as shown in
Fig~\ref{fig_detects}.  Region B extends 3.5\degr\ along the vector
line and includes all cluster members detected, as well as some non-members 
outside the 3.5\degr\ cluster radius.

After Local Group (LG) correction, the mean velocity of galaxies is
936 $\pm$ 63 \kms in region A, 1266 $\pm$ 68 \kms in region B and 1459
$\pm$ 62 \kms in region C.  The Kolmogorov-Smirnov (KS) statistic 
indicates that the mean
LG-corrected velocities of galaxies in regions A and C differ by 523
$\pm$ 125 \kms, significant at the 99.99 per cent  confidence level.
Comparing the mean peak flux in the two regions, we find it to be
lower in region C than in A by up to $\sim$55 per cent , significant at the
95 per cent  confidence level. The mean total flux of the two regions is not
significantly different. From these comparisons we conclude that a
significant gradient in galaxy velocity is measureable across the
field.

This gradient closely follows the velocity gradient of optical 
galaxies evident in Plate 8 of the 
Nearby Galaxies Atlas, showing the Southern Supercluster in the region of 
Fornax (Tully \& Fisher, 1987).

The radial velocity of the central cD galaxy of the Fornax cluster,
NGC\,1399, is 1425 km\,s$^{-1}$.  We have measured the recessional velocity of
the three galaxies for which Cepheid distances have been determined in
the \textit{Hubble Space Telescope} Key Project on the Extragalactic
Distance Scale (Kennicutt, Freedman \& Mould 1995).  Madore et al.\ (1999)
determined the distance to NGC\,1365 to be 18.6 $\pm$ 0.6 Mpc. We
measure a recessional velocity of 1638 \kms, consistent with the
suggestion of Drinkwater et al.\ (2000) that the galaxy is in front
of NGC\,1399 and is falling towards the cluster centre.  Prosser et
al.\ (1999) estimated the distance to NGC\,1326A to be
18.7 $\pm$ 2.0 Mpc.  The galaxy is part of the subcluster identified by
Drinkwater et al.\ (2001a) and identified by the small ellipse in
Fig~\ref{fig_dens}.  Our measured velocity of 1836 \kms supports their
conclusion that the subcluster is in front of the main cluster and
falling inwards. The Cepheid distance to NGC\,1425 was calculated to
be 22.2 $\pm$ 2.7\Mpc\ (Mould et al.\ 2000).  The large-scale structure
we have detected suggests that NGC\,1425 may be at a greater distance
than the Fornax cluster proper. From its velocity of 1512 \kms and
location 5.6\degr\ north of the cluster, we would expect it to be
about 1.5 Mpc more distant than the cluster centre.

\subsection{\hi masses and mass-to-light ratios}  %% ~\ref{subsec_M/L}
\label{subsec_M/L}

The \hi masses of detections are plotted in Fig.~\ref{fig_HIMF} with
the highest detection rate being in the 10$^{8.5}$--10$^{9.0}$\msun\,
bin, i.e., close to the survey detection limit.  For our whole sample,
estimated \hi mass-to-blue-light ratios ($L_B$ uncorrected for
inclination) range from 0.1 to $\sim$11, with a mean of 1.7,
although for a small number of detections the uncertainty in optical
identification requires further clarification.  The range of \hi mass
and \ML values as a function of distance from the centre of Fornax are
plotted in Figs.~\ref{fig_MHI_R} and ~\ref{fig_ML_R},
respectively. The values appear relatively independent of distance to
the cluster centre.  However, these two plots have ignored possible
projection biases. For example, our highest \hi mass cluster member is the
bright spiral NGC\,1365 (J0333-36), circled in Fig.~\ref{fig_MHI_R}.  There is
evidence that this galaxy is in front of the cluster and falling
towards it (Drinkwater et al. 2001a) and therefore its \hi mass may be
lower than our estimate and its actual  distance from the
cluster centre greater than the projected distance measured here.
This figure indicates that almost 80 per cent  of our detections have \hi
masses below $2\times10^9$ $M_\odot$ and that most of the higher mass
galaxies are outside the cluster region.

The distribution of \ML values plotted as a function of projected
distance to the cluster centre is indicated in Fig.~\ref{fig_ML_R}.
The circled detection, FCC\,302, has a low redshift ($cz = 803$ km\,s$^{-1}$),
a high \ML value ($\sim$5.2) and may be behind the cluster and
infalling.  

The only \hi detection seen in projection within the core
region is FCC\,235 (NGC\,1427A), indicated by the triangle in
Figs.~\ref{fig_MHI_R} and ~\ref{fig_ML_R}. Its irregular morphology,
particularly high redshift ($cz = 2024$ km\,s$^{-1}$) and \ML value of
$\sim$1.3 support the arguments of Hilker et al.\ (1997), Chanam\'{e}, 
Infante \& Reisenegger (2000) and Drinkwater, Gregg \& Colless (2001a) that 
the galaxy is in
front of the cluster centre and infalling.  If these data points are
ignored, our results are suggestive of reduced \ML values and \HI depletion 
within 4\degr\ of the cluster centre.  We measure a cluster mean \ML value of
1.2 $\pm$ 0.3 and a mean of 1.7 $\pm$ 0.2 for galaxies more than 4\degr\
from the centre, the difference being significant at the $\sim$90 per cent  level. 
We note here that standard measures of \HI deficiency do not include 
non-detections and interpretation of such results remains thereby 
inherently uncertain.

The deeper, targeted \hi observations of Schr\"{o}der did, however,
detect significantly lower \ML in Fornax cluster galaxies
(0.68 $\pm$ 0.15) compared to a selected field sample (1.15 $\pm$ 0.10)
(Schr\"{o}der et al.\ 2001). This is probably explained in part by
their more sensitive observations and larger sample of cluster
galaxies.  In addition, Schr\"{o}der et al.\ also found that the
cluster galaxies show significant \hi depletion as measured by the
Solanes parameter (Solanes, Giovanelli \& Haynes 1996).  Although \hi depletion
measurements were beyond the scope of our observations, our \hi
measurements are consistent with Schr\"{o}der's results for the 23
common galaxies observed.

In Fig.~\ref{fig_M/L_M}, the distribution of \ML and \mhi values is
plotted.  Of detections with \ML$>1$,  $\sim$80 per cent 
are fairly low \mhi galaxies with \mhi$<2\times10^9$ \msun.
These galaxies are also all of low luminosity.  Although corrections
for inclination have not been made here, we conclude that most of the high
\ML detections are small or dwarf galaxies. Again, projection effects
have not been considered in this plot.

%----------- fig 12 -----------
\begin{figure}
\psfig{file=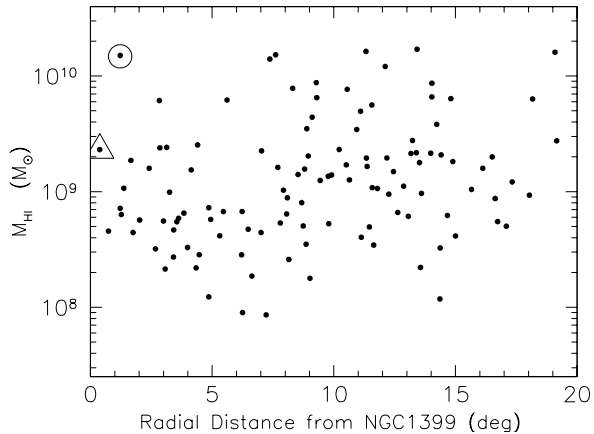,angle=-90,width=80mm}
\centering
\caption{More than 50 per cent  of detections have an \hi mass less than
$10^9$ M$_\odot$.  Most of the high \hi mass detections are outside
the cluster region.  The circled point, NGC\,1365, and NGC1427A,
identified by the triangle, are discussed further in Section
~\ref{subsec_M/L}.}
\label{fig_MHI_R}
\end{figure}

%----------- fig 13  -----------
\begin{figure}
\centering
\psfig{file=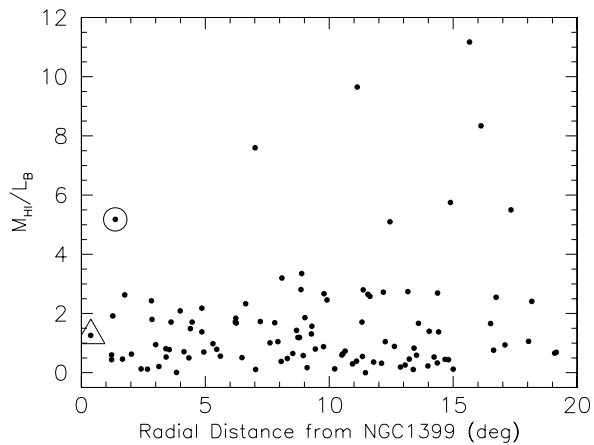,angle=-90,width=80mm}
\caption{The \ML ratios of detections, as a function of projected
distance from the cluster centre, are fairly evenly distributed across
the mosaic region.  Almost 50 per cent  have a \ML ratio $>$1
{\mbox{M$_\odot$/L$_\odot$}}.  The circled detection is J0345-35
(FCC302) which has a high \ML (5.2) and a low redshift ($cz = 803$ 
km\,s$^{-1}$). The measured radial offset may thus be biased by a projection
effect. The galaxy may be beyond the cluster and infalling. The object
J0340-35 (FCC235) (identified by the triangle) has a very high
redshift of 2028 \kms and is probably in front of the cluster centre
and infalling. } 

\label{fig_ML_R}
\end{figure}

%----------- fig 14  -----------
\begin{figure}
\psfig{file=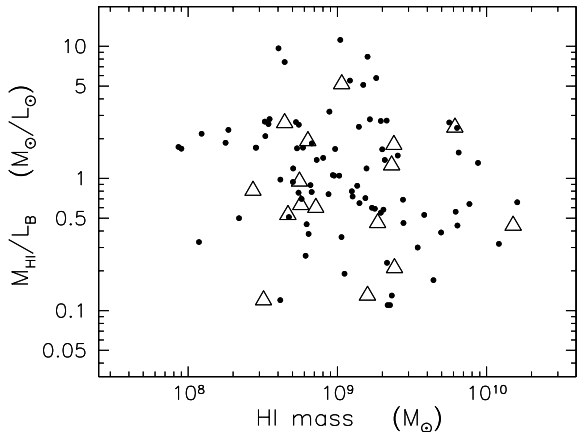,angle=-90,width=80mm}
\centering
\caption{In this plot of \mhi versus \ML, cluster detections are shown
as triangles and detections outside the cluster are shown as solid
dots.  The majority of detections ($\sim$84 per cent ) with \ML $>$1 are small
or dwarf galaxies with low luminosity and low \hi masses
($<2\times10^9$ \msun).}
\label{fig_M/L_M}
\end{figure}

%----------- section CONCLUSION -----------
\section{Conclusions}
\label{sec_concl}

We have mosaicked an area of $\sim$620 \sqd around the
Fornax cluster of galaxies using standard HIPASS cubes. An intensive
search of this cube has resulted in a catalogue of 110 galaxies. Of
these, 14 galaxies have newly determined redshifts, 4 have velocity
measurements significantly different to values published for the
likely optical counterpart and for about 10 per cent  the optical
identification is currently uncertain.  This \hi catalogue can be used
in studies of the large-scale structure around the Fornax cluster, the
morphological evolution of galaxies and the morphology-density
relation in the cluster and field environments.  \\

Our main results are:

\begin{enumerate}
\item A strong indication that cluster centres which are readily 
apparent in the counts of optical 
galaxies are only detected as \hi overdensities of a few times the density of 
{\hbox{H{\sc{i}}}}-detected galaxies in the field. 

\item The region of overdensity of \hi galaxies extends for several
degrees ($\sim$1.7 Mpc) around the cluster, including
the cluster infall region.

\item The large-scale \hi structure around the Fornax cluster is
confined to a plane with similar velocity width to velocity dispersion
we measure for the cluster itself (425 $\pm$ 71 km\,s$^{-1}$). This plane is
almost perpendicular to the line-of-sight but we measure a significant
difference of $\sim$500 \kms in the redshifts of galaxies in the
NW region compared to the SE region.  Further analysis of the LSS will
enable peculiar velocities to be decoupled from the general
cosmological expansion.
\end{enumerate}

Published optical magnitudes have been used to estimate \ML ratios.  From 
this we can conclude that:

\begin{enumerate}
 
\item There is no unambiguous signal of \hi depletion in the small number of
{\hbox{H{\sc{i}}}}-detected galaxies near the cluster core. 

\item Only a small fraction of galaxies are detected in \hi near the 
cluster core.  Morphologically, the \hi-detected galaxies are all 
late-type and these are less likely to be found in regions of high 
galactic density.

\item The galaxies with higher \ML are all dwarf-type galaxies.

\end{enumerate}

Further work intended in these studies includes the resolution of
`confused' sources by deeper \hi imaging at the ATCA (Australia Telescope 
Compact Array). 

Many questions about galaxy evolutionary processes in and around
clusters remain unanswered.  Our studies indicate that the core of the
Fornax cluster is deficient in \hirich galaxies, a finding which is
tightly correlated with the morphologies of galaxies in the core.
However, the evolution of galaxies in the cluster environment is
currently poorly understood and future work will be aimed at further
elucidation of these processes.

\vspace{1.0cm}

\noindent\textbf{NOTES FOR TABLE 1.}\\

\noindent Explanatory notes for the columns in Table 1 are given in this 
subsection.  (All values are from unsmoothed spectra.)\\

\noindent {\textit{Col. (1)}} HIPASS name (Jhhmm-dd). Suffixes a, b, etc. are used to name uniquely HIPASS detections when required. \\ 
{\textit{Cols. (2)}} and {\textit{(3)}} \hi
detection center position in R.A. and dec. (J2000). \\ 
{\textit{Cols. (4)}} and {\textit{(5)}}
\hi Heliocentric systemic radial velocity in \kms (optical convention)
at the midpoint of the 50  and 20 per cent  widths ($V_{50}$ and $V_{20}$)
 respectively, measured from unsmoothed spectra. \\
{\textit{Cols. (6)}} and {\textit{(7)}} The 50   and 20 per cent  \hi velocity widths in \kms ($W_{50}$ and $W_{20}$). \\ 
{\textit{Col.  (8)}} \hi peak flux in Jy.  \\ 
{\textit{Col. (9)}} Total integrated \hi flux density in Jy\,\kms.  \\ 
{\textit{Col. (10)}} Apparent blue
magnitude (from NED, measurement type shown in Col. (11)).\\ 
{\textit{Col. (11)}}
Type of magnitude measurement in Col. (10). \\ 
{\textit{Col. (12)}} \hi
mass-to-blue light ratio (\ML). \\ 
{\textit{Col. (13)}} Optical galaxy velocity
(km\,s$^{-1}$) recorded in NED. \\ 
{\textit{Col. (14)}} Projected separation in
arcminutes (\textit{d}\arcmin) of the positions of the \hi detection and the
likely corresponding optical galaxy. \\ 
{\textit{Col. (15)}} Name(s) of the
likely matching optical galaxy. \\ 
{\textit{Col. (16)}} Remarks --- including the
names(s) of possible optical galaxy match(es) where the match is
uncertain. These detections are labelled `C' (`confused'). 
Derived values shown in the table proper refer to the galaxy named in 
Col. (14).  For detections labelled `V diff', the \HI velocity differs 
significantly from the published optical value(s). The comment `new V' 
indicates that a velocity measurement has not previously been published.\\

\section{Acknowledgements}
M.W.  gratefully acknowledges the support of the University of
Melbourne in the provision of a Melbourne Research Scholarship.
We are most appreciative of the anonymous referee's helpful comments which have 
resulted in valuable improvements to this work.

%-----------  BIBLIOGRAPHY  -----------

\label{lastpage}

%----------- section appendix/HI spectra -----------
\appendix
\section{\hi spectra}

The \hi spectra of all detections are plotted in the Appendix. The
profiles show heliocentric radial velocity (optical convention) 
in \kms versus flux in Jy.
Spectra shown here have been Hanning-smoothed and a Gaussian-smoothed 
baseline fitted over the velocity range shown.
Several spectra have been masked before flux measurements were
made --- the detection being measured is indicated with an arrow and
the regions masked are enclosed in dotted lines (see \emph{Notes}
below).
Dashed lines on the spectra show the velocity range within
which $V_{50}$ and $V_{20}$ were determined, with the pairs of dots on
the spectra delineating the velocity widths, $W_{50}$ and $W_{20}$, at
50  and 20 per cent  of the peak flux.
Flux and velocity width measurements
were made on unsmoothed spectra and so may vary slightly from values
suggested by the smoothed spectra.

\section{Notes on Selected Spectra}
\begin{description}

\item {\textit{J0308-23}} The \HI emission at $\sim$2880 \kms is from
IC1892. The extra emission suggested at a slightly higher velocity but
below the 3$\sigma$ limits, is possibly from MCG-04-08-029 with a NED
velocity of 3016 km\,s$^{-1}$. 

\item {\textit{J0322-37}} Nearby emission at $\sim$1200--1350 \kms has
been masked before fitting the baseline to this spectrum. 

\item {\textit{J0323-36}} The central (arrowed) emission in this spectrum
is from NGC1326. Emissions from the nearby interacting galaxies
J0325-36a (NGC1326A, 1836 km\,s$^{-1}$) and J0325-36b (NGC1326B, 1006 km\,s$^{-1}$)
are also seen in this spectrum but have been masked out in
measurements. 

\item {\textit{J0325-36a}} Emission from J0325-36b (NGC1326B) is also seen
in this spectrum. 

\item {\textit{J0325-36b}} Emission from J0325-36a (NGC1326A) is also seen
in this spectrum.  

\item {\textit{J0335-32}} This detection is at the lowest peak flux of the 
            observations described in this work. We have listed it as
            `confused' since the spectrum shows considerable
            baseline ripple, the positional discrepancy with the
            optical coordinates is greater (4.8 arcmin) than for most
            confirmed detections and our velocity measurement differs
            significantly from published optical values (see, for
            example, de Vaucouleurs et al., 1991).

\item {\textit{J0341-22}} Considerable ripple can be seen in this spectrum
and if the fitted baseline shown is subtracted, the profile more
obviously has the typical double-horned shape of an inclined spiral
galaxy (NGC1415). The velocity and velocity width measurements are in
close agreement with the values published by Andreani, Casoli \& Gerin 
(1995).

\item {\textit{J0359-45}} The main emission in this spectrum is from the
irregular galaxy known as the Horologium Dwarf.  Also seen at a
slightly higher velocity is emission from the nearby edge on spiral
ESO249-G035 which has a published (NED) velocity of 1031 \kms but is
below the flux cutoff levels of this survey.

\end{description}

\bsp

%----------- table -----------

\newpage

\begin{figure*}
\centering  
\vspace{-20mm}
\psfig{file=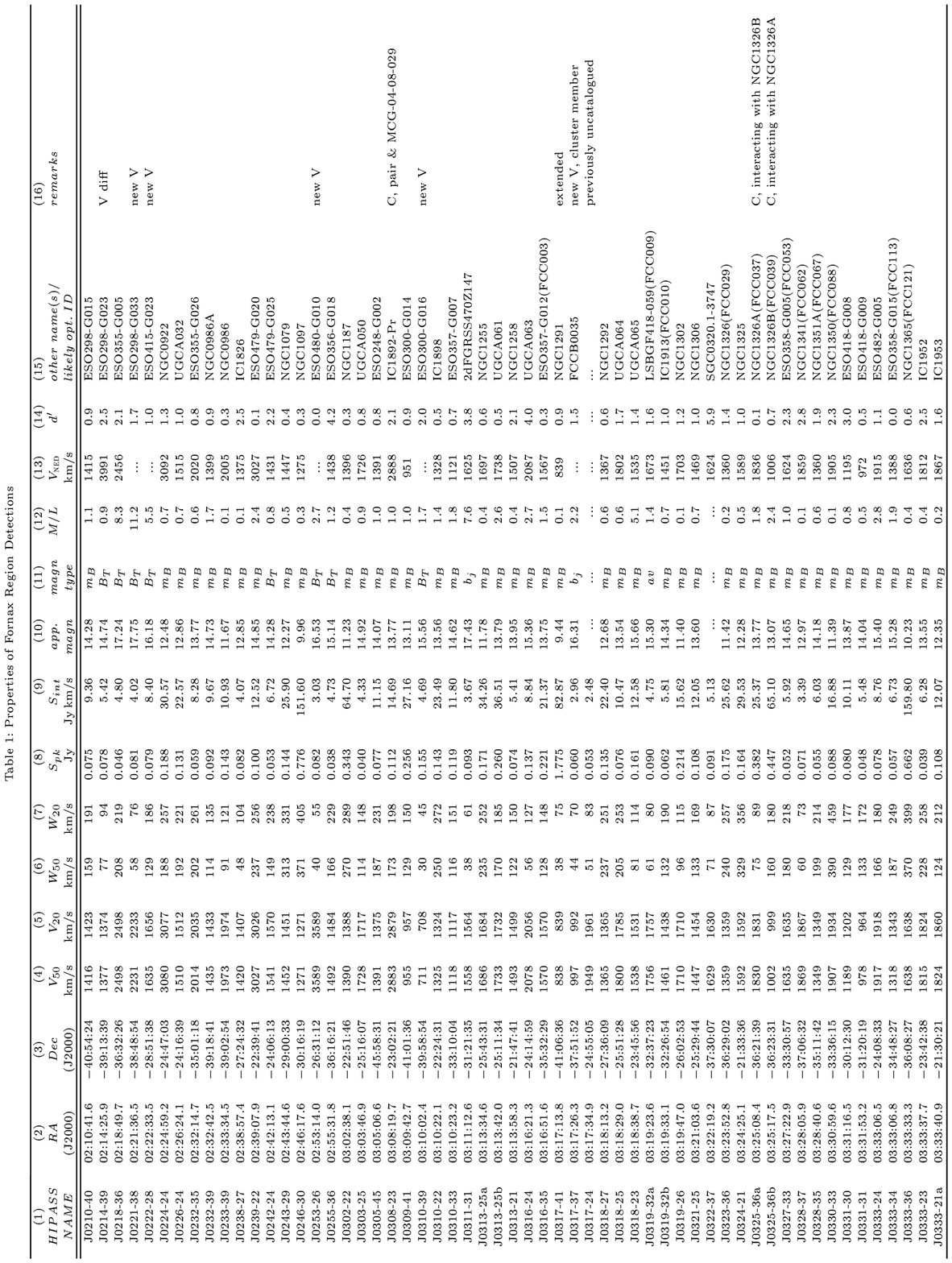,angle=0,width=200mm}
\label{tab1a}
\end{figure*}

\newpage

\begin{figure*}
\centering  
\vspace{-20mm}
\psfig{file=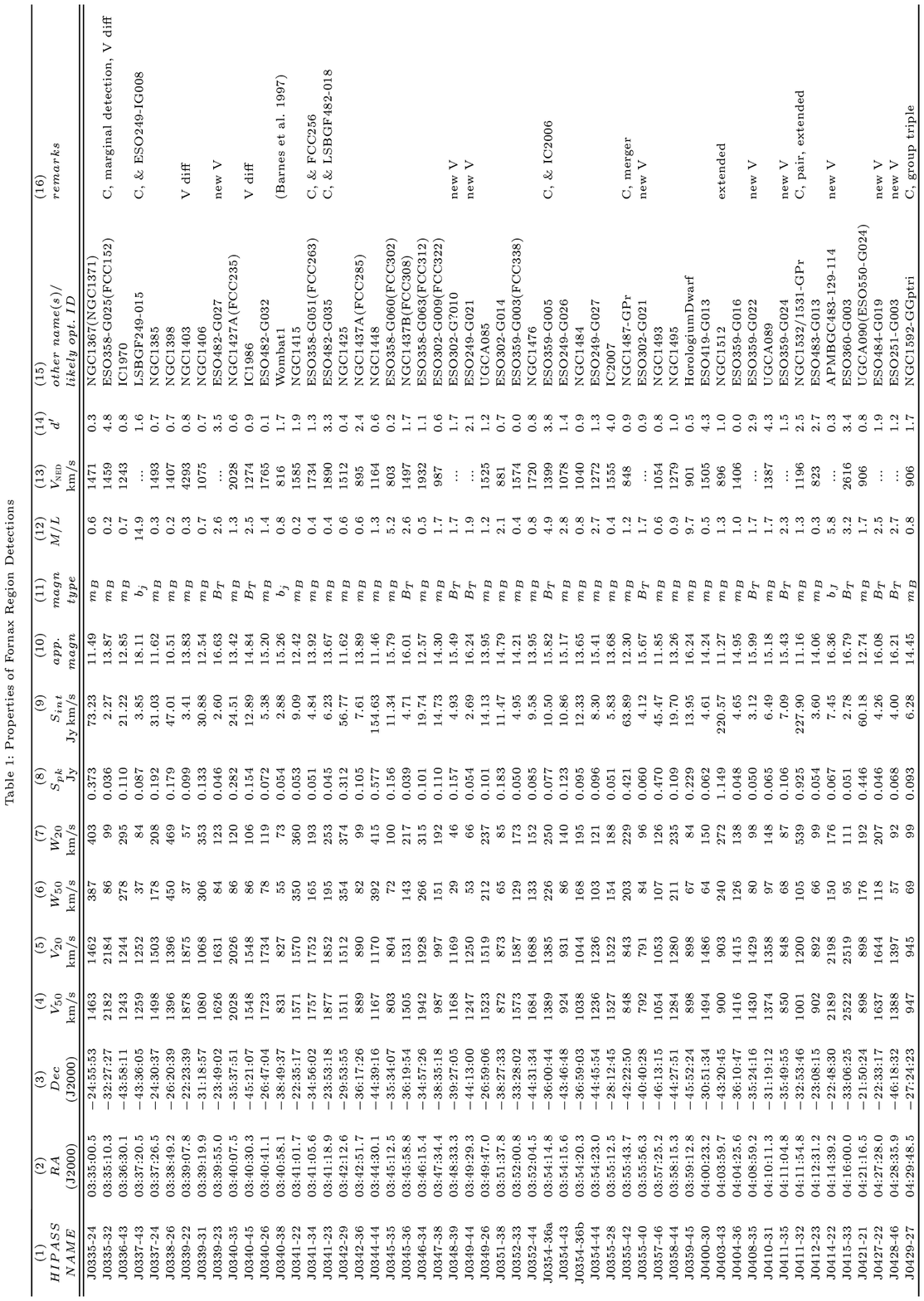,angle=0,width=200mm}
\label{tab1b}
\end{figure*}

% \newpage

\begin{figure*}   
\centering                                                          
\label{spect1}         
\end{figure*} 

\begin{figure*}                                                              
\centering                                                                  
\label{spect2}         
\end{figure*}

\begin{figure*} 
\centering                                                 
\centering                                                                  
\label{spect3}         
\end{figure*}

\begin{figure*}
\centering  
\centering                                                                  
\label{spect4}         
\end{figure*}                                                                
 
\end{document}